\documentclass[
 aip,
 amsmath,amssymb,
 reprint,%
]{revtex4-1}

\usepackage{graphicx}
\usepackage{dcolumn}
\usepackage{bm}

\usepackage[utf8]{inputenc}
\usepackage[T1]{fontenc}
\usepackage{mathptmx}
\usepackage{etoolbox}
\usepackage{xcolor}

\makeatletter
\def\@email#1#2{%
 \endgroup
 \patchcmd{\titleblock@produce}
  {\frontmatter@RRAPformat}
  {\frontmatter@RRAPformat{\produce@RRAP{*#1\href{mailto:#2}{#2}}}\frontmatter@RRAPformat}
  {}{}
}%
\makeatother
\begin{document}

\preprint{AIP/123-QED}

\title[Hole-Based Stealthy Hyperuniform Semiconductor Metamaterials for the Mid-Infrared]{Hole-Based Stealthy Hyperuniform Semiconductor \\Metamaterials for the Mid-Infrared}
\author{Manuel Gallego}%
 \email{mg6693@princeton.edu}
\affiliation{ 
Department of Electrical and Computer Engineering, Princeton University, Princeton, NJ 08544, USA
}%
\author{Sara Kacmoli}%
\affiliation{ 
Department of Electrical and Computer Engineering, Princeton University, Princeton, NJ 08544, USA
}%
\affiliation{ 
Omenn-Darling Bioengineering Institute, Princeton University, Princeton, NJ 08544, USA
}%
\author{Yezhezi Zhang}%
\affiliation{ 
Department of Electrical and Computer Engineering, Princeton University, Princeton, NJ 08544, USA
}%

\affiliation{ 
Current address: Cruise LLC, 333 Brannan St, San Francisco, CA 94107, USA
}%

\author{Michael A.~Klatt}%
\affiliation{ 
German Aerospace Center (DLR), Institute for AI Safety and Security, Wilhelm‐Runge‐Str.~10, 89081 Ulm, Germany
}%
\affiliation{ 
German Aerospace Center (DLR), Institute for Material Physics in Space, 51170 Köln, Germany
}%
\affiliation{ 
Department of Physics, Ludwig-Maximilians-Universität, Schellingstr.~4, 80799 Munich, Germany
}%
\author{Claire F.~Gmachl}%
\affiliation{ 
Department of Electrical and Computer Engineering, Princeton University, Princeton, NJ 08544, USA
}%

\date{\today}

\begin{abstract}
Stealthy hyperuniform heterostructures are a novel type of metamaterial with the potential for optical image processing at angles away from normal incidence. These metamaterials show analogous properties to photonic crystals while circumventing the spatial anisotropy often hindering the latter's use. In this paper, we have successfully designed, fabricated, and characterized a hole-based stealthy hyperuniform structure on a quantum cascade layer substrate. The infrared spectral data reveal a sizable gap-midgap ratio of 9.8\% for a photonic band gap around $12.0 ~\mu m$ in the form of an enhanced reflection region for increasing incidence angles. The stealthy hyperuniform metamaterial also showed spatial isotropy by its unchanging reflection spectrum for all in-plane rotational angle measurements.
\end{abstract}

\maketitle

\section{Introduction}
Disordered hyperuniform structures can form a novel type of metamaterial with inherent hidden order\cite{torquato2003local}. Hyperuniform point distributions were originally discovered by Torquato and Stillinger in a systematic characterization of local density fluctuations\cite{torquato2003local}. These point distributions can create an exotic state of matter that combines properties from both liquids and crystals due to the hidden order within the disordered structure\cite{torquato_hyperuniform_2018,zhang2022novel,florescu2013optical, leseur2016high,man2013photonic, dal2022waves, vynck2023light, aubry2020experimental, froufe2017band}; most notably, recent work indicated that so-called stealthy hyperuniform heterostructures are the only ones that can possess a complete photonic band gap (PBG) in the themodynamic limit\cite{torquato_hyperuniform_2018,florescu_designer_2009, klatt2022wave} (number of points grows to infinity). Initially, PBGs have been thought to be exclusive to ordered structures such as photonic crystals and quasicrystals~\cite{joannopoulos_photonic_2008, man2005experimental}; however, through a powerful collective-coordinate optimization technique \cite{klatt2022wave, torquato2015ensemble, uche2004constraints, batten2008classical,florescu2013optical}, it is possible to engineer a stealthy hyperuniform structure with targeted structure factor, with a PBG at a particular wavelength, and angle range~\cite{gorsky_engineered_2019,yu_engineered_2021}.

Stealthy hyperuniform point patterns \cite{batten2008classical,torquato2015ensemble} are a subclass of hyperuniform point distributions, in which the structure factor (an order parameter) vanishes below a critical wave vector, $k_c$. This is equivalent to vanishing density fluctuations in the infinite wavelength limit\cite{florescu_designer_2009}.
Furthermore, the disordered nature of the pattern makes stealthy hyperuniform devices spatially isotropic while retaining a PBG, as opposed to their photonic crystal counterparts that also possess PBGs but are highly anisotropic from their spatial symmetries. These unique properties of stealthy hyperuniform structures have enabled novel applications in optics and photonics such as the fabrication of disordered silicon photonic waveguides \cite{milovsevic2019hyperuniform}, mode selection in THz quantum cascade lasers (QCLs) \cite{degl2016hyperuniform}, tailored light scattering \cite{piechulla2021tailored}, disordered lenses\cite{zhang2019experimental}, potential computational metamaterials for edge detection \cite{zhang2022novel,guo2018photonic}, among others\cite{zhang2021hyperuniform,tavakoli2022over,merkel2024stealthy,wan_hyperuniform_2023}. Other potential applications of stealthy hyperuniform devices are to create in-plane resonators for QCLs by replacing one of the facets with a disordered structure.

In this work we demonstrate a stealthy hyperuniform pattern on QC layered semiconductor substrates by using etched holes. The size of the pattern unit cell is 30 $\mu m$ x 30 $\mu m$  with average center to center point distance of 3 $\mu m$; this unit cell is then tiled across a 1 $cm^2$ sample. The designed PBG is located at 12.3 $\mu m$, and we indeed observe the PBG at 12.0 $\mu m$ using a broadband IR source. We also observe the spatial isotropy of the stealthy hyperuniform pattern by rotating it about its own plane at a fixed incidence angle.

\begin{figure*}
    \centering\includegraphics[width=14cm]{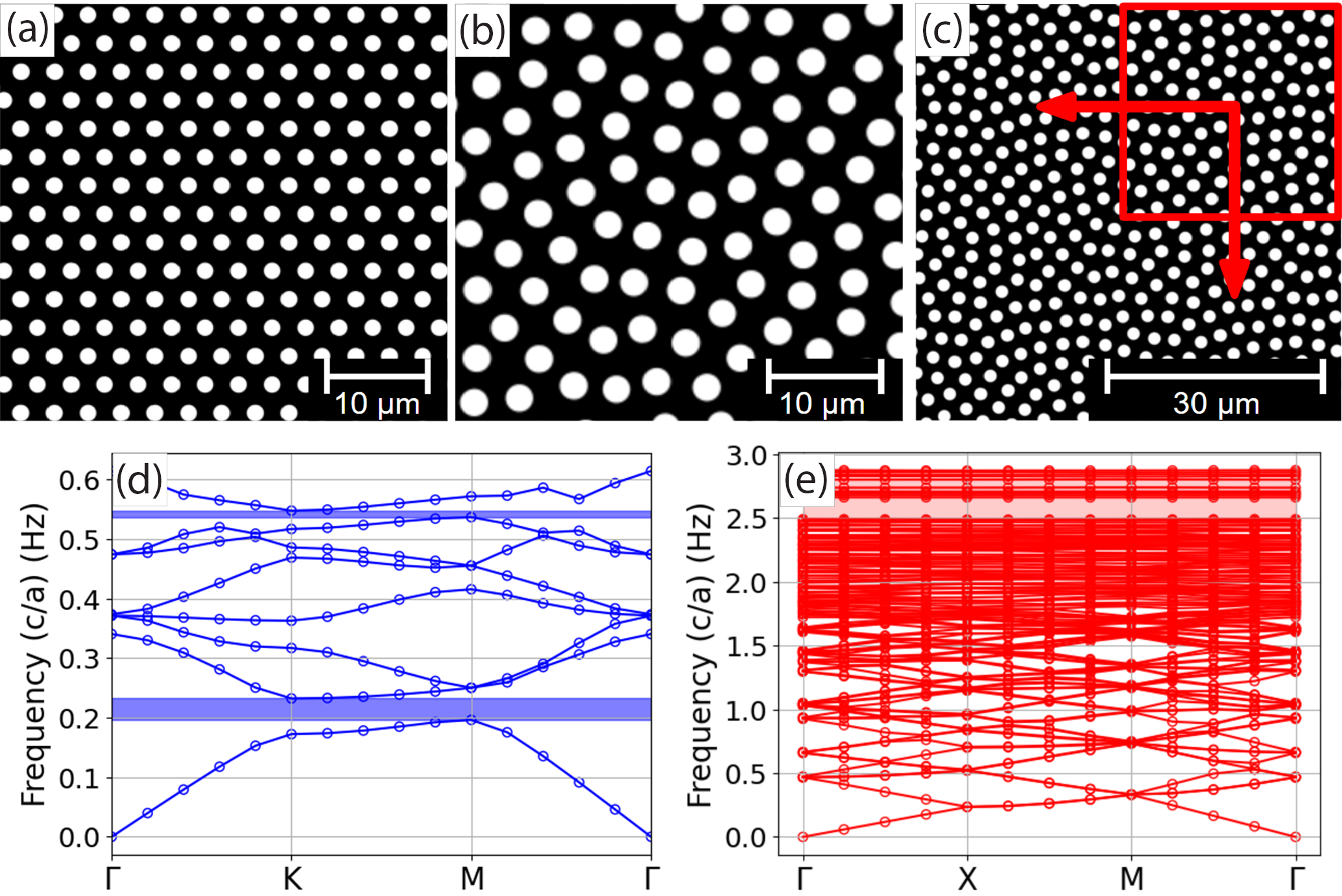}
    \caption{(a) Hexagonal photonic crystal hole pattern with lattice constant $a=3$ $\mu m$ and hole diameter of 1.5 $\mu m$, (b) one unit cell of the stealthy hyperuniform hole pattern with hole diameter of 2.2 $ \mu m$ and average center to center hole spacing of 3 $ \mu m$ , and (c) The stealthy hyperuniform unit cell in (b) tiled into a square lattice. (d) TE band diagram of the hexagonal hole photonic crystal in (a). (e) TE band diagrams of the stealthy hyperuniform hole pattern in (c). Around 110 individual bands were calculated to observe the stealthy hyperuniform pattern band gap at around 2.4 (c/a) Hz, i.e. 12.3 $\mu m$ wavelength. $c$ is the speed of light in vacuum, and $a=30~\mu m$ is the size of the unit cell.  }
    \label{sim}
\end{figure*}

\section{Photonic Band Gap Simulation}
A point distribution is \textit{hyperuniform} if the variance of the number of points in an observation window of radius $R$ grows asymptotically slower than the area of the window itself. This definition equates to a structure factor $S(k)$ of zero for $k\rightarrow0$, where $k$ is the wavevector in reciprocal space\cite{torquato2003local,torquato_hyperuniform_2018}. 

We here use a stealthy hyperuniform point pattern from Florescu et al 2009 and compute its photonic band structure with the electromagnetic mode solver (MPB)~\cite{johnson2001block}, which uses harmonic mode solutions to Maxwell's equations. Our hole-based stealthy hyperuniform semiconductor metamaterial was modeled as a periodic geometry consisting of air holes with infinite depth in a material with dielectric constant of $\varepsilon = 10.6$ (Fig.~\ref{sim}b,~\ref{sim}c). Similarly, a triangular photonic crystal (Fig.~\ref{sim}a,~\ref{sim}d) with lattice constant $a = 3~\mu m$ was modeled for comparison. The Stealthy hyperuniform pattern was tiled as a square lattice to fill the entire computational domain, so the Brillouin zone by consequence is that of the square lattice. The mode solver operates by computing the associated frequencies and bands along the edges of the Brillouin zone, allowing us to construct band diagrams highlighting the existence, or lack thereof, of a PBG (Fig.~\ref{sim}e).  Since our stealthy hyperuniform unit cell is rather small (102 points) we compute over the entire domain to highlight the spatial isotropy (flat bands).

We computed the band diagrams, in both transverse electric (TE) and transverse magnetic (TM) polarizations. In the case of our stealthy hyperuniform hole geometry, the PBG appears in TE polarization but not in TM (if the geometry were to be inverted to dielectric pillars in air, then the PBG would appear in the TM polarization but not in TE \cite{zhang2022novel}). For a stealthy hyperuniform pattern with 2.2 $\mu m$ diameter holes, the TE band gap was computed to be centered at approximately 12.3 $\mu m$ (810 $cm^{-1}$) with 7.0\% width. We note that the bands become flatter at higher frequencies from the statistical isotropy of the pattern. The diameter of the holes was optimized by selecting the largest gap-midgap ratio from a range (0.5 - 3 $\mu m$) of diameters. The minimum and maximum ranges are constrained by the smallest possible fabrication requirements, and the overlapping of holes, respectively. Hole diameters from 1.8 - 2.2 $\mu m$ were found to be best, leading to a 5 - 7\% wide PBG. 

\begin{figure*}
    \centering\includegraphics[width=15.5cm]{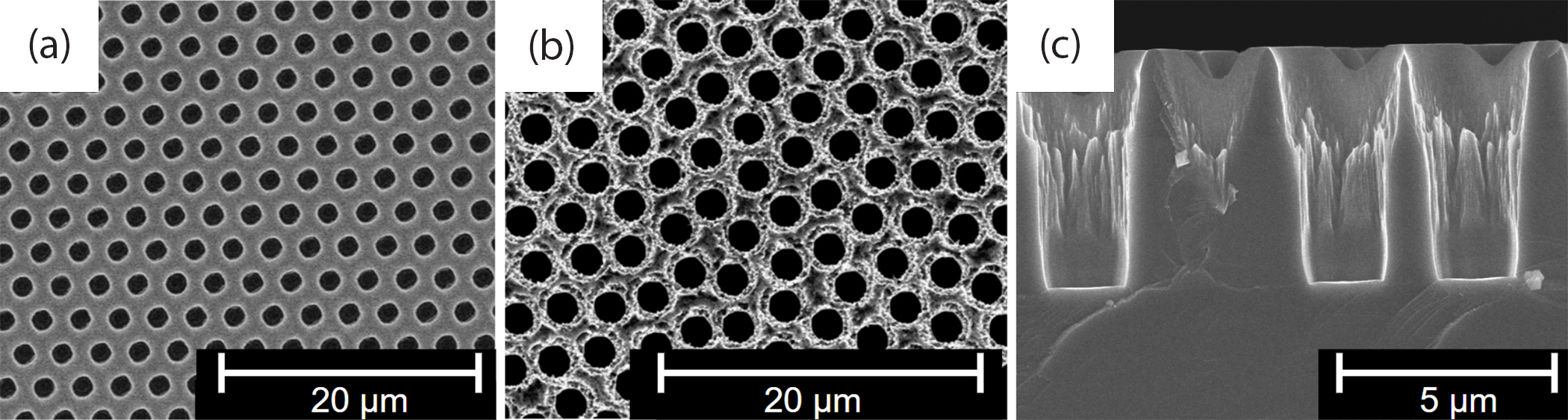}
    \caption{Scanning electron microscopy (SEM) images of (a) the top view of the hexagonal photonic crystal, (b) the top view of the stealthy hyperuniform sample with hole diameter of $1.8~\mu m$, and (c) and the cross section of the stealthy hyperuniform sample. The surface roughness visible in (b) and (c) is primarily due to mask erosion from the etch step. However this is not expected to affect measurements since the roughness is significantly smaller than the wavelengths used.}
    \label{fab}
\end{figure*}
\section{Device Fabrication}
The stealthy hyperuniform samples consist of the same pattern shown in Fig.~\ref{sim}b tiled across a 1 $cm^2$ substrate. The hole diameters of the samples we fabricate vary from 0.9 - 2.1 $\mu m$, and 3 $\mu m$ average distance between centers. In addition, an unpatterned sample, and a hexagonal photonic crystal with 3 $\mu m$ lattice constant and 1.5 $\mu m$ hole diameter were also fabricated for reference (Fig.~\ref{fab}a,~\ref{fab}b). All samples are fabricated on an InP substrate with epitaxially-grown QC layers. 
The device fabrication involves standard optical photolithography and reactive ion etching (RIE). The deep etching of the semiconductor layers is performed using Cl$_2$ and BCl$_3$ chemistry at high temperatures (220$^\circ$ C)~\cite{kacmoli2022unidirectional}. For this reason, we use an SiO$_2$  hard mask patterned via direct-write lithography as opposed to a more common photoresist mask. Our process is developed to achieve smooth and vertical sidewalls both in the hard mask and the  semiconductor itself. The holes are etched deep through the active core of the QC device and into the bottom cladding region (Fig.~\ref{fab}c). Lastly, we cleave the sample, and lap and polish it down to about 250 $\mu m$ thick for increased transmission. To prepare for measurements, the samples are mounted to a copper block and placed in a custom-built rotational mount.

\section{Experimental setup}
To conduct a thorough study of transmission and reflection spectra, a setup outside of the Fourier Transform Infrared (FTIR) Spectrometer is created. It consists of the globar source inside the FTIR, an external cryogenically cooled mercury cadmium telluride (LN$_2$-MCT) photodetector, three rotational stages to adjust sample and detector angles, and a polarizer. The collimated light from the FTIR first passes through a set of focusing ZnSe lenses, and then through a wire grid polarizer to select between TE or TM modes. The sample is mounted on a 3D printed rotational mount and placed at the focal point of the focusing lenses. The rotational mount  can vary the incidence angle on the sample surface, as well as rotate about its own in-plane axis for in-plane rotational angle measurements. The set-up is designed to work in both transmission and reflection mode allowing rays away from normal incidence to pass unobstructed. This is important because the effect of the PBG is more evident at larger incidence angles. 
\onecolumngrid

\begin{figure}[!b]
\centering
\includegraphics[width=17cm]{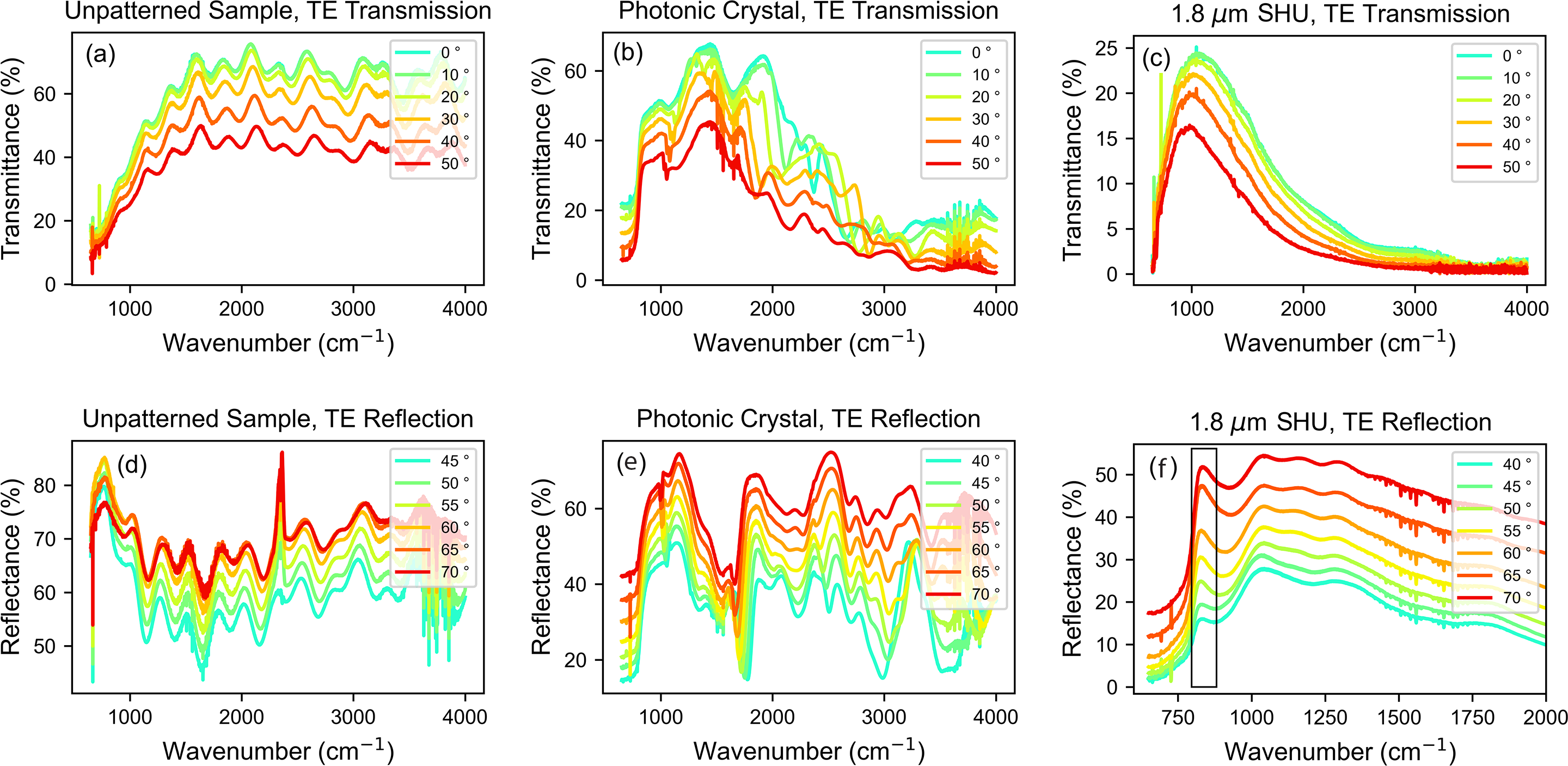}
\caption{ FTIR spectra of (a,d) the unpatterned sample with QC epilayers, (b,e) the photonic crystal sample with lattice constant $a=3~\mu m$ and hole diameter of $1.5~\mu m$, and (c,f) the $1.8~\mu m$ stealthy hyperuniform (SHU) sample at varying incidence angles as indicated. The transmittance (a,b,c) and reflectance (d,e,f) signals follow the expected behavior of Fresnel TE coefficients with increasing incidence angle. The unpatterned sample shows oscillations from thin film interference, while maintaining an almost identical spectral shape for all incidence angles. The spectral shape of the photonic crystal varies significantly with increasing incidence angles due to out of plane change of the crystal structure. The 1.8 $\mu m$ stealthy hyperuniform sample exhibits a PBG as an enhanced reflection region around 835 $cm^{-1}$ marked in (f) as a boxed region. Both transmittance and reflectance SHU plots retain their spectral shape in the entire range of incidence angles.}
\label{spectraA}
\end{figure}
\twocolumngrid 
\textcolor{white}{.}

\textcolor{white}{.}

The photodetector used in this set up is a LN$_2$-MCT detector that rotates around the sample on a swiveling arm. Two apertures are mounted in front of the photodetector to reduce oversampling, followed by another ZnSe lens to focus the light onto the surface of the photodetector.
\begin{figure}[b]
    \centering
    \includegraphics[width=6.1cm]{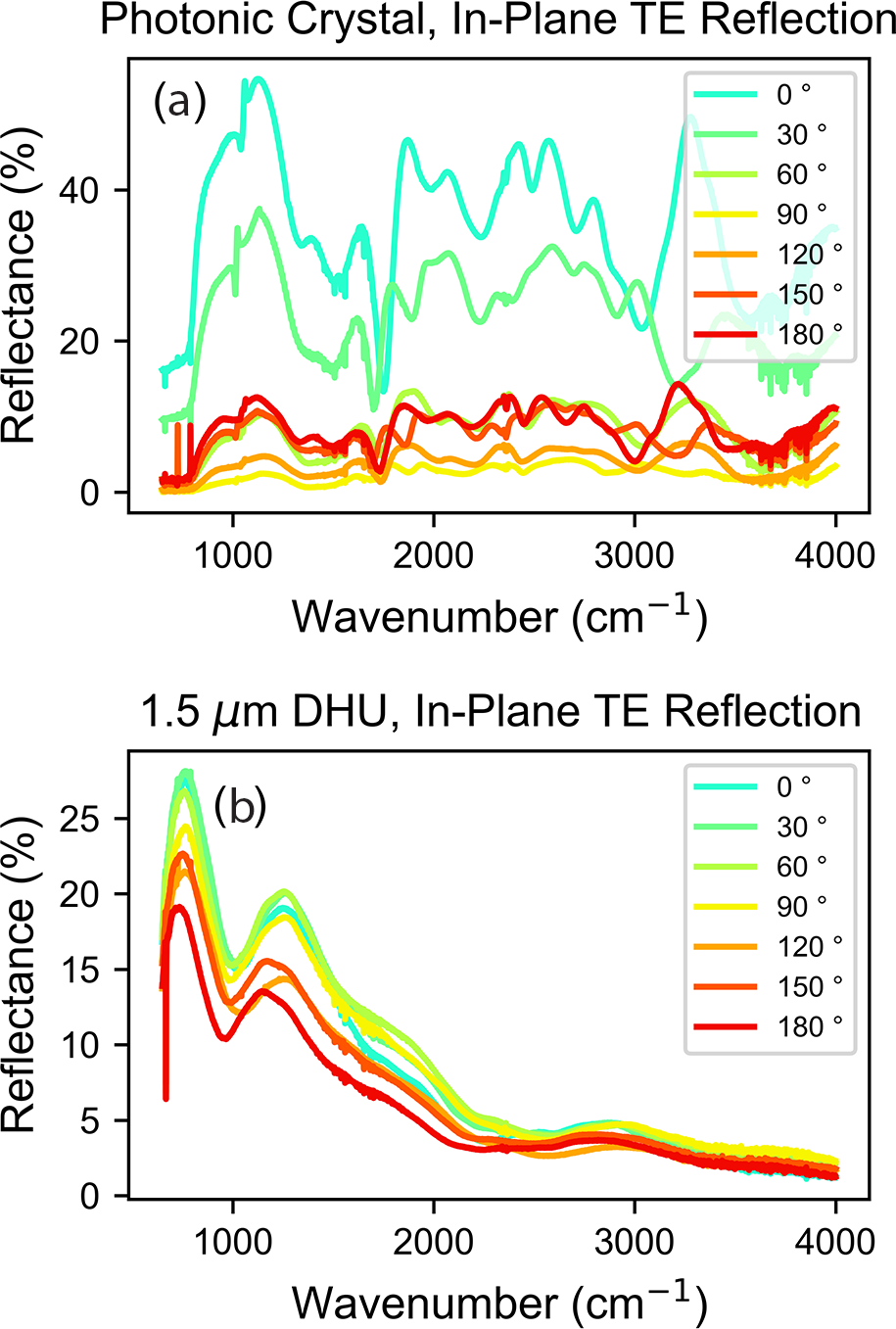}
    \caption{FTIR spectra of (a) the photonic crystal, and (b) the $1.5~\mu m$ stealthy hyperuniform sample showing the in-plane reflection of the samples as a function of mount rotation angle at a fixed incidence angle of 45$^\circ$. The photonic crystal has a different spectral shape at various in-plane rotational angles while the 1.5 $\mu m$ stealthy hyperuniform spectral shape remains the same. This is a clear demonstration of spatial anisotropy from the photonic crystal's three fold symmetry and the spatial isotropy of the stealthy hyperuniform structure. }
    \label{spectra B}
\end{figure}
\section{Infrared Spectrum Analysis}
A background spectrum was recorded for all measurements to eliminate atmospheric features, such as H$_2$O and CO$_2$ absorption peaks, from the sample spectra. The overall signal strength of all sample measurements follow the Fresnel TE transmission and reflection coefficients, so transmittance and reflectance decrease and increase, respectively, as the angle of incidence grows.

The unpatterened sample spectrum serves as a reference spectrum to compare the stealthy hyperuniform and photonic crystal samples; it contains oscillations at various incidence angles in the transmission and reflection spectra consistent with thin film interference from the QC epilayers (Fig.~\ref{spectraA}a, ~\ref{spectraA}d). 

The photonic crystal sample displays decreasing transmittance with increasing wavenumbers, while the reflectance spectrum has prominent peaks throughout the entire spectral range. One can also see the strongly varying spectral features with increasing incidence angles for both transmission and reflection (Fig.~\ref{spectraA}b,~\ref{spectraA}e.) which is expected from the out of plane change of the crystal structure.
Notably, its largest calculated PBG is around 670 $cm^{-1}$ lies outside of the detection range of the detector so its not visible in the spectrum.

The 1.8 $\mu m$ stealthy hyperuniform sample shows an increased transmission signal at small wavenumbers (500 $cm^{-1}$ - 1500 $cm^{-1}$) with decreases at  growing wavenumbers in both transmission and reflection modes; the vanishing signal at large wavenumbers is due to light scattering from the holes (Fig.~\ref{spectraA}c, ~\ref{spectraA}f). The overall smooth spectral shape is similar for all incidence angles except for where the PBG is located. Reflection spectra  show a distinct enhanced reflection region around 835 $cm^{-1}$ (about 12.0 $\mu m$) corresponding to the stealthy hyperuniform PBG (Fig.~\ref{spectraA}f). The measured PBG is close to the predicted one at 810 $cm^{-1}$; the 3.1\% error can be attributed to the imperfect hole surface and patterning, as well as the finite and infinite hole depth from the sample and the simulation, respectively. The width of the experimentally measured PBG was computed to be 9.8\% from Gaussian fitting; bigger than the 7.0\% PBG width from simulation. 

Furthermore, we tested if our finite stealthy hyperuniform structures do, in fact, possess spatial isotropy. The samples were set at a fixed incidence angle of 45$^\circ$ and then rotated from 0$^\circ$ to 180$^\circ$ about their own plane; the in-plane rotational angle spectra for the photonic crystal and the stealthy hyperuniform structure are shown in Fig.~\ref{spectra B}a and Fig.~\ref{spectra B}b, respectively. The stealthy hyperuniform sample conserves its spectral shape throughout the entire range of measured in-plane rotation angles which purely comes from the spatial isotropy. On the other hand, the in-plane rotational angle spectra of the photonic crystal sample show the three-axis symmetry of the hexagonal (or triangular) crystal lattice; most notably measurements in discrete increments of 60$^\circ$ (e.g. 0$^\circ$, 60$^\circ$, 120$^\circ$, 180$^\circ$) have matching spectral features. 

\section{Conclusion}
In this study we have designed, fabricated, and characterized stealthy hyperuniform patterned devices for the mid-IR region, and compared them to a reference unpatterned sample as well as a photonic crystal structure. Under TE reflection for the stealthy hyperuniform sample, a noticeable band gap around 835 $cm^{-1}$ ($\lambda = 12.0 ~\mu m$) was detected in the form of an enhanced reflection region with 9.8\% PBG width.
Furthermore, the stealthy hyperuniform in-plane rotational angle measurements showed spatial isotropy from similar spectral shapes, resulting in a in-plane isotropic PBG.

A future application of this stealthy hyperuniform pattern may be to become a functional computational metamaterial capable of performing edge detection in the mid-IR range by making use of its designed PBG. 

\section{Acknowledgments}
We would like to thank the Micro and Nano Fabrication Center (MNFC) at Princeton University where the devices were fabricated. 

We acknowledge Salvatore Torquato for valuable discussions.

M.A.K. acknowledges funding and support by the Deutsche Forschungsgemeinschaft (DFG, German Research Foundation) through the SPP2265, under grant numbers KL 3391/2-2, WI 5527/1-1 and LO 418/25-1, and by the Initiative and Networking Fund of the Helmholtz Association through the Project ``DataMat''.

\section*{Data Availability Statement}
The data that support the findings of this study are available from the corresponding author upon reasonable request. 

The following article has been submitted to APL Photonics journal. After it is published, it will be found at [link pending]
\section{References}
\bibliography{aipsamp}

\end{document}